\documentclass[prd,aps,
tightenlines,
superscriptaddress,showpacs,nofootinbib,eqsecnum,amsfonts,amsmath
,twocolumn
]{revtex4}
\usepackage{bm,graphics,graphicx,epsfig,color
}

\def\be{\begin{equation}}
\def\ee{\end{equation}}

\def\gsim{\mathrel{
\rlap{\raise 0.511ex \hbox{$>$}}{\lower 0.511ex
\hbox{$\sim$}}}}
\def\lsim{\mathrel{
\rlap{\raise 0.511ex \hbox{$<$}}{\lower 0.511ex
\hbox{$\sim$}}}}

\begin{document}
\title{Precessing supermassive black hole binaries and dark energy measurements with LISA}

\author{Adamantios Stavridis} \email{astavrid@physics.wustl.edu} 
\affiliation{McDonnell Center for the Space Sciences, Department of
Physics, Washington University, St.  Louis, Missouri 63130, USA}

\author{K. G. Arun} \email{arun@physics.wustl.edu} 
\affiliation{McDonnell Center for the Space Sciences, Department of
Physics, Washington University, St.  Louis, Missouri 63130, USA}
\author{Clifford M. Will} \email{cmw@wuphys.wustl.edu} 
\affiliation{McDonnell Center for the Space Sciences, Department of
Physics, Washington University, St.  Louis, Missouri 63130, USA}
\affiliation{GReCO, Institut d'Astrophysique de Paris, CNRS,\\ 
Universit\'e Pierre et Marie Curie, 98 bis Boulevard Arago, 75014 Paris, France}
\noaffiliation{}

\begin{abstract}
Spin induced precessional modulations of gravitational wave signals 
from supermassive black hole binaries can improve the estimation of
luminosity distance to the source by space based gravitational wave 
missions like the Laser Interferometer Space Antenna (LISA). 
We study how this impacts the ablity of LISA  to do cosmology, specifically, to measure
the dark energy equation of state (EOS) parameter $w$. 
Using the $\Lambda$CDM model of cosmology, we show that observations of 
precessing binaries with mass ratio 10:1 by LISA, combined with a redshift measurement, 
can improve the determination of $w$ up to an
order of magnitude with respect to the non precessing case
depending on the total mass and the redshift.
\end{abstract}

\date{\today}
\maketitle

\section{Introduction}
When the proposed orbiting Laser Interferometer Space Antenna (LISA) detects an inspiralling compact
binary system, it can not only localize the source on the sky but can also
measure its luminosity distance independent of astronomical distance ladder calibrations.
If an electromagnetic (EM) counterpart associated with this
GW event provides the redshift to the source, then the combination
of these observations can have profound cosmological implications~
\cite{Schutz86}, such as precise determinations of Hubble's
constant~\cite{Schutz86,Markovic93,ChernoffFinn93,MacLeodHogan07,SSV09,CutlerHolz09,Nissanke09a}
and measurements of the dark energy equation of state
parameter $w$~\cite{HolzHugh05,DaHHJ06}.


Spin effects may also help to improve cosmological measurements using gravitational waves.
If the compact binary components have non-aligned spins, then the
modulations induced by precession~\cite{ACST94,K95} 
can break degeneracies between various parameters being estimated and improve accuracy,
especially of the sky location and luminosity distance~\cite{Vecchio04,LangHughes06,LangHughes07}.
Recently we have developed a code to carry out parameter estimation for precessing 
inspiralling massive binary black holes, using the Fisher matrix formalism, 
in order to consider the impact of spin precession on LISA's 
ability to distinguish a general class of massive theories of graviton
from general relativity~\cite{SW09} (see also a similar work by \cite{YagiTanaka09}).  
In this report, we use a variant of this code to study  the cosmological 
implications of the improved distance measurements possible with spinning 
massive black hole binaries.
 
Vecchio~\cite{Vecchio04} first pointed out the possible improvements in precision 
provided by precessions induced by a subset of spin-orbit couplings.  
Lang and Hughes~\cite{LangHughes06,LangHughes07} generalized this to the full 
panoply of spin orbit as well as spin-spin effects.  
Ref.~\cite{LangHughes06} focused on the improvement
in the estimation of masses and spins of the binary and briefly discussed 
improvements in distance measurements. Their follow-up paper~\cite{LangHughes07}
showed that precessing binaries would offer much better angular localization by LISA 
and discussed how electromagnetic follow-ups could be used effectively to identify 
the host galaxy and obtain the redshift (see also Ref.~\cite{LangHughes08}).

\begin{figure}[t]
\includegraphics[width=2.8in]{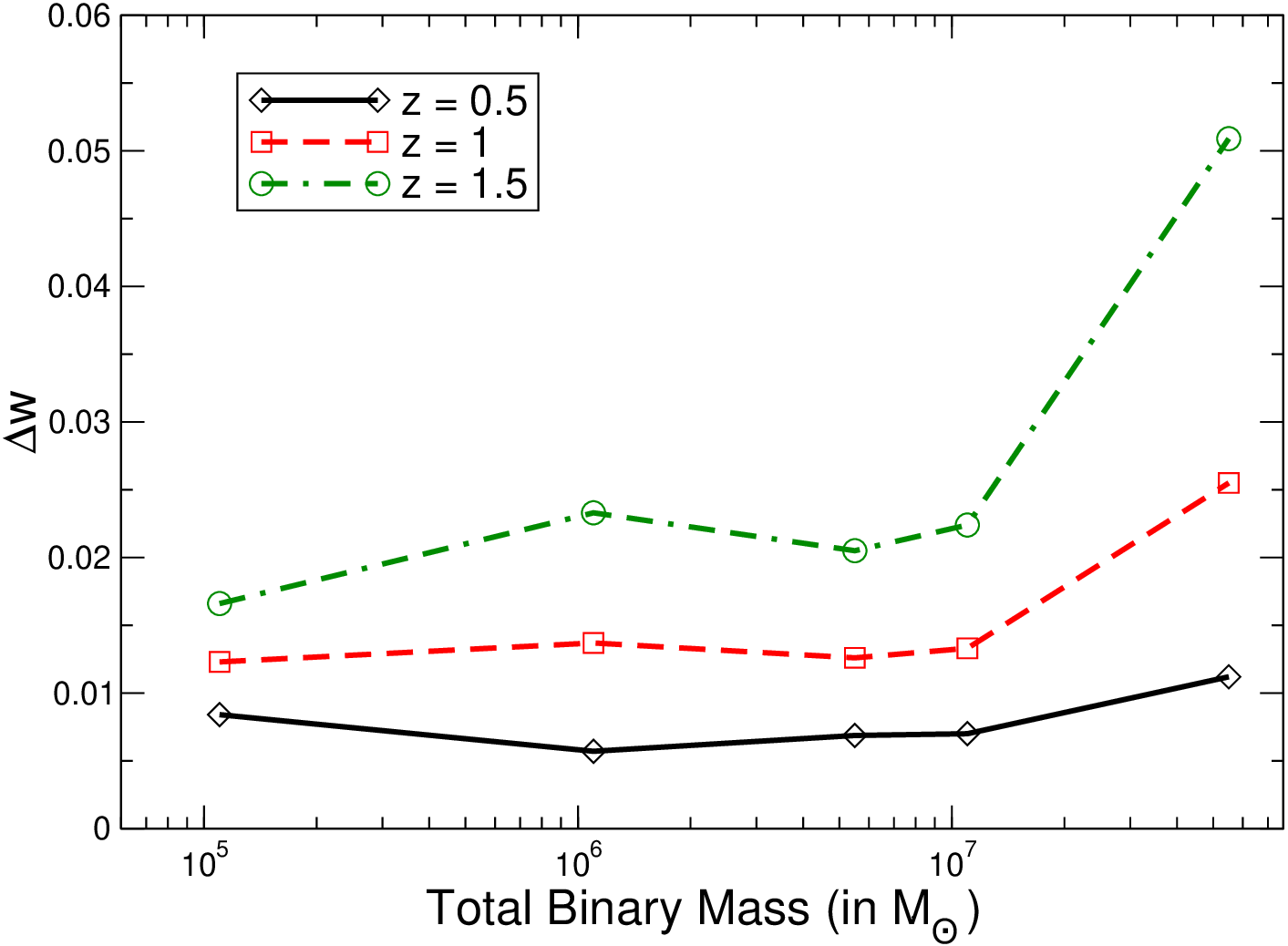}
\caption{$1\sigma $ errors in the dark energy EOS parameter 
$w$ as a function of the total binary black hole mass at redshifts 
of $0.5$, $1$ and $1.5$. Binaries are all assumed to have precession and contain black holes of mass
$(1,10)\times 10^4$, $(1,10)\times 10^5$, $(5,50) \times 10^5$, $(1,10) \times 10^6$, 
$(5,50) \times 10^6$ $M_\odot$. The data points are medians of
$10^4$ runs for each mass and redshift, and ignore the effect of weak lensing.}
\label{Fig:Mass_z}
\end{figure}

In this paper, we show explicitly that improved distance measurements with precessing binaries 
combined with a redshift to the source could lead to precise measurements of $w$; the results 
are summarized in Fig. \ref{Fig:Mass_z}.   For example, for a binary system of 
$(1+10) \times 10^6 M_\odot$ at redshift $z=1.5$, the median $1\sigma$ error in measuring $w$, 
over an ensemble of $10^4$ binaries distributed randomly in spins, orbital orientations, 
and sky locations is about 2 percent.

\section{Models and assumptions}

Our waveform model is described in detail in~\cite{SW09}. 
We use second-post-Newtonian (2PN) accurate ``restricted'' waveforms (RWF) for binaries on quasi-circular orbits
and include the precession effects of spin-orbit and spin-spin coupling.
The stationary phase approximation is used for computing  the Fourier transform of the signal. 
Though spin contributions at 2.5PN~\cite{BBuF06} and nonspinning terms up to 3.5PN order ~\cite{BDEI04} 
are available for the RWF, we assume that they induce only corrections to the leading effects studied here. 
However, incorporation of spin dependent higher harmonics~\cite{K95,ABFO08} could have some influence on 
the results~\cite{KleinEtAl09}.
We follow Cutler~\cite{Cutler98} in our model of the LISA satellite and its orbital
motion. The noise characteristics of LISA that we assume and use are the same
as in Ref.~\cite{BBW05a} and used by Lang and Hughes~\cite{LangHughes07}.
We have taken into account the effect of precession on the antenna pattern functions of LISA.

We use a Fisher matrix analysis to estimate the errors in estimating the 15 parameters that 
characterize the system: two masses, two dimensionless spin magnitudes (which vary from 0 to 1), 
the time and phase of coalescence, four sets of two angles each specifying the location of the binary, 
the initial angular momentum direction and the two initial spin directions of the binary's members 
(eight in total), and finally the luminosity distance. All the angles used are with respect to the 
solar system barycenter. In the specific case that the individual spin vectors of the black holes are 
aligned, (a rather optimistic case astrophysically) the two extra sets of angles (four parameters) 
for the individual spins are not needed since the binaries do not precess~\cite{BBW05a}.
We also assume that LISA provides two independent signal outputs with uncorrelated noises.  
Finally, we assume that the sources are observed for one year prior to coalescence.

For a given choice of the physical masses of the two black holes and of the redshift 
or luminosity distance, we distribute $10^4$ sources randomly in the sky with random 
values of the remaining 10 parameters (we choose coalescence time and phase to be one year and 
zero, respectively, in all cases).  For each of the realizations, we solve numerically the precessing 
equations during the inspiral phase of the system, compute the output signals $h^{\rm I},h^{\rm II}$, 
and the Fisher information matrix defined as, 
\begin{equation}
\Gamma_{ab} \equiv \left( \frac{\partial h^{\rm I}}{\partial\theta^a} \mid
\frac{\partial h^{\rm I}}{\partial\theta^b} \right) + 
\left( \frac{\partial h^{\rm II}}{\partial\theta^a} \mid
\frac{\partial h^{\rm II}}{\partial\theta^b} \right) \, ,
\label{Eq:FisherMatrix}
\end{equation}
where the inner product is,  
\begin{equation}
(h_1|h_2)
\equiv  4 \, {\rm Re} \int_0^{\infty} df \frac{\tilde{h}_1^* (f)  \tilde{h}_2 (f)}{S_n(f)} \, ,
\end{equation}
where $\tilde{h}(f)$
denotes the Fourier transform of the gravitational waveform
$h(t,\theta^a)$, star denotes complex conjugate, $S_n(f)$ is the noise
spectral density of the detector, and $\partial h/\partial \theta^a$ denotes
the partial derivative with respect to the parameter $\theta^a$ being
estimated.  The superscripts $I$ and $II$ denote the two LISA outputs.  The
signal-to-noise ratio (SNR) for a given signal $h(t,\theta^a)$ is then given
by,
$\rho[h] \equiv  (\tilde{h}|\tilde{h})^{1/2}$.  We then invert the Fisher
matrix to obtain the covariance matrix $\Sigma^{ab}$, and the corresponding
root mean square errors from the square roots of the diagonal entries, as follows,
\begin{equation}
\Delta\theta^a = \sqrt{\Sigma^{aa}} \,, \qquad  \Sigma = \Gamma^{-1} \,.
\label{errors}
\end{equation}
We focus on systems with a mass ratio of $10:1$ since these are astrophysically interesting and 
exhibit stronger effects of spin precession.

\begin{figure}[t]
\includegraphics[width=2.8in]{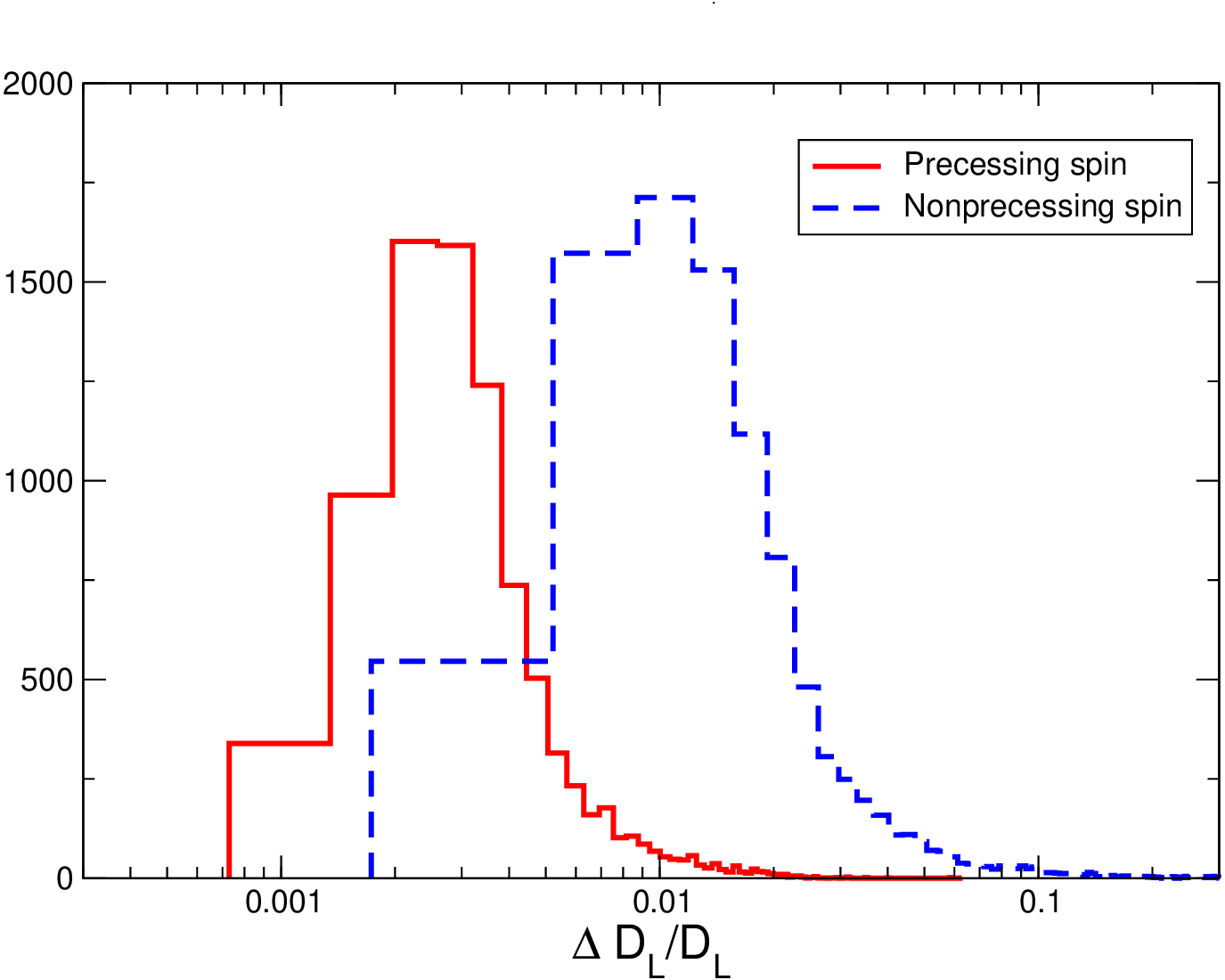}
\caption{Distribution of the $1\sigma$ errors in luminosity distance measurement based on $10^4$ 
random realizations. 
The mass of the binary system is $(10^5+10^6)M_\odot$ at a redshift of $z=1$. 
Solid (red) histogram is for precessing systems; dashed (blue) histogram is for systems with spins aligned.}
\label{Fig:Lumdistz1}
\end{figure}

Fig.~\ref{Fig:Lumdistz1} shows the histogram of the relative errors in the luminosity distance
for a system with masses $(10^5,10^6) \,M_\odot$ at a redshift of $z=1$. 
Precession of the binary has a significant impact on the distribution as the peak 
of the distribution shifts to the left (improves) almost by an order of magnitude 
with respect to the case where all systems in the population are nonprecessing. 
This is in reasonable agreement with the results of Lang and Hughes~\cite{LangHughes07} 
(see, e.g., figure 7 of their paper).

However, to do cosmology, we need the redshift of an  electromagnetic signal associated with
an observed LISA  event or with its host galaxy. 
To date there is no clear understanding of the astrophysical mechanisms that could
produce an electromagnetic afterglow (or a precursor) to a binary black hole merger, 
although a number of possibilities have been discussed~\cite{KFHM06}. 
Whether LISA can localize the source on the sky so that
extensive electromagnetic follow up missions can be launched 
has been addressed in many recent works.  Of particular interest
here are the estimates of Lang and Hughes~\cite{LangHughes07},
showing that typically for a $z=1$ source of total mass $10^6 M_\odot$, 
the angular resolution taking precession into account is about $0.3-3$ square degrees
one day prior to merger (see Table 5 of \cite{LangHughes07} for example).
Kocsis et al~\cite{KHMtrig08} argued that finding EM counterparts associated with LISA sources 
will be difficult, but may be achievable with the advent of various
wide field telescopes which would be operational by the time LISA flies 
(see Table I of Ref.~\cite{KHMtrig08} for details). Further, 
some authors have argued~\cite{HolzHugh05,KFHM06} that the number of candidate galaxies associated 
with a LISA event, can be reduced by 2-3 orders of magnitude
by looking for the source within a 3D error volume, combining the angular resolution of LISA and 
the approximate luminosity distance LISA would provide in advance. 
For this reason (rather optimistically) we have not rejected any distance estimate in the $10^4$ 
realizations based on the size of the angular error box or the detectability of an EM counterpart.

\begin{figure*}[t]
[htp]
\begin{center}
\begin{tabular}{c}
\includegraphics[width=8cm]{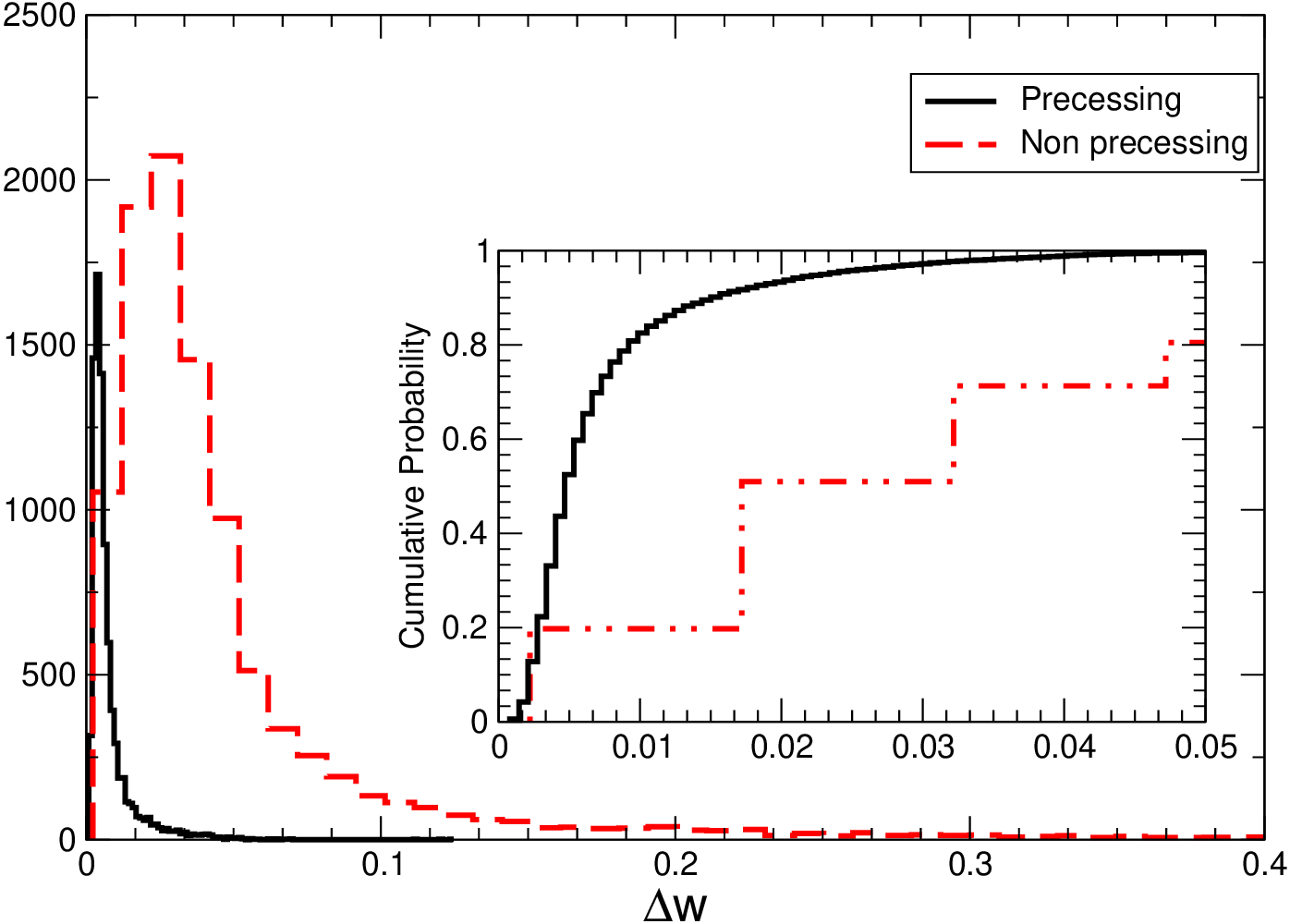}
\hskip 1.0 true cm
\includegraphics[width=8cm]{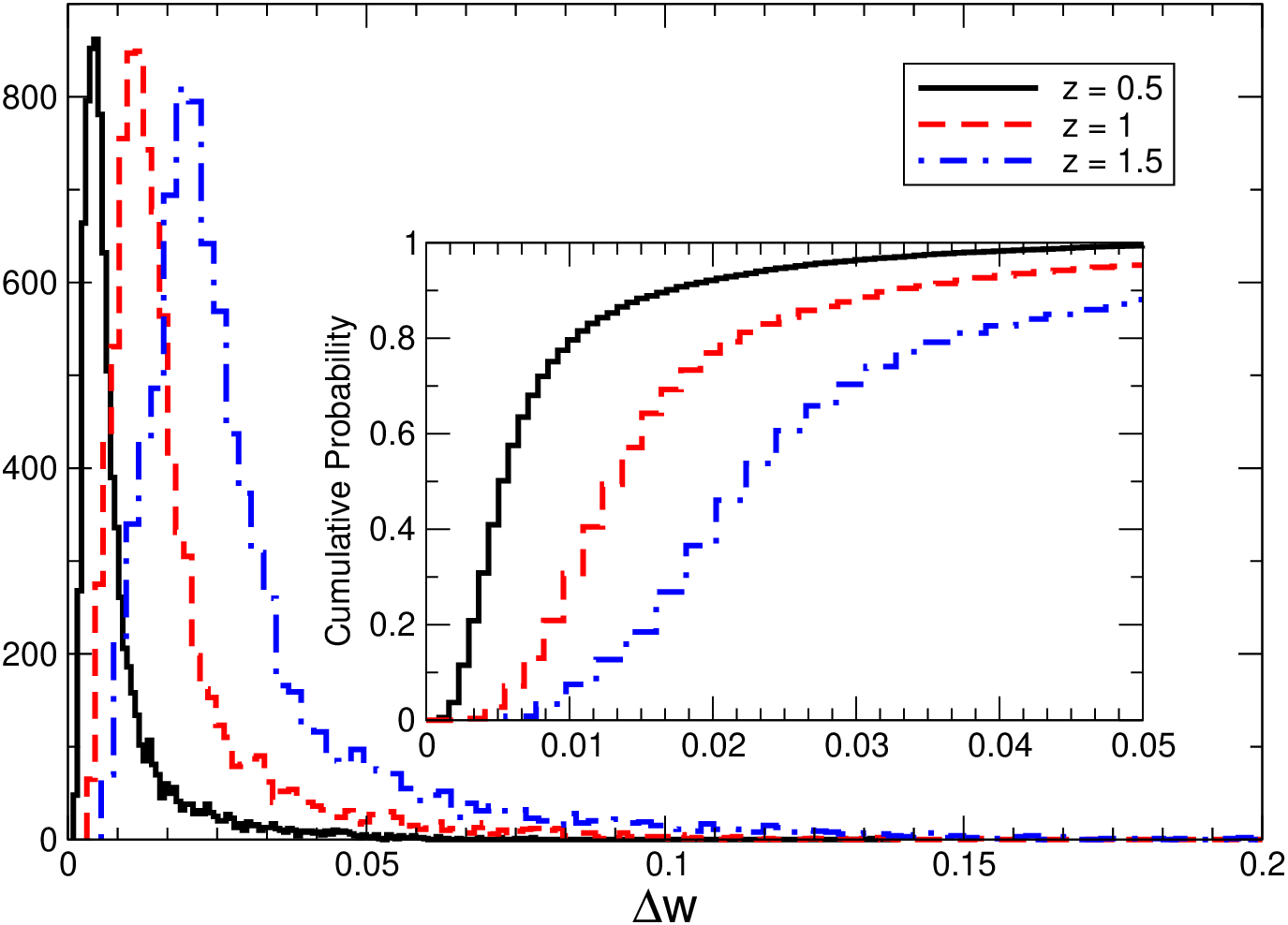}
\end{tabular}
\end{center}
\caption{Distributions of $1\sigma$ errors in $w$ for various values of redshifts and spin configurations. 
Left panel: Distribution of the errors in the dark energy EOS parameter $w$ for a binary system
of masses $10^5+10^6\, M_\odot$ at redshift $z=1$, and for the precessing and non-precessing cases. 
The inset frame shows a zoom of the corresponding cumulative histograms.
Right panel: Distribution of errors in $w$ for the same binary system with masses $10^5+10^6 \, M_\odot$ 
in the precessing case for three different redshifts. 
The inset frame shows a zoom of the corresponding cumulative histograms.}
\label{Fig:Dwerrors}
\end{figure*}
To translate distance errors into errors in the dark energy EOS parameter $w$, 
we work with the standard cosmological model with a flat universe and the nominal parameters
$H_0 = 75\, {\rm km \, s^{-1} \, Mpc^{-1}}$, matter density $\Omega_M = 0.25$,
dark energy density $\Omega_{\Lambda}=0.75$ and $w=-1$.
The luminosity distance of a source at a redshift $z$ is given by,
\begin{equation}
D_L = (1+z) \int_0^z { dz'\over H(z') } \, ,
\label{Lumdist}
\end{equation}
where $H(z)$ is given by
\begin{equation}
H(z) = H_0 \sqrt{ \Omega_M (1+z)^3 + \Omega_{\Lambda}(1+z)^{3(1+w)}} \,.
\label{Hubblez}
\end{equation}

In order to illustrate directly the effect of precession, we assume 
that errors in $H_0$, $\Omega_M$, $\Omega_\Lambda$ and $z$ are negligible, so that
the error in $w$ is related directly to the error in $D_L$ by~\cite{AISSV07}
\begin{equation}
\Delta w = D_L \left| \frac{\partial D_L}{\partial w} \right|^{-1} \frac{\Delta D_L}{D_L}\,.
\label{Dw}
\end{equation}
The value of ${\partial D_L}/{\partial w}$ can be calculated using (\ref{Lumdist}) and (\ref{Hubblez}) 
for the different values of the redshift used. 

Finally, we have neglected the effect of weak lensing on the estimation of the
luminosity distance. This is a serious caveat,  especially
at redshifts above 1. However, mechanisms have been proposed which might help to
reduce the weak lensing error in the future.  These include ``corrected
lenses"~\cite{Dalal03WL},  the accurate measurement of the weak lensing power spectrum
with radio observations~\cite{Pen04WL}, and combining optical and infra-red
observations of foreground galaxies~\cite{Goobar06}. 
See~\cite{ShapiroEtAl09} for a recent discussion.


\section{Results and Discussion}%
The inclusion of spin precession and its attendant modulation of the GW waveform has a 
significant impact on the accuracy of measurements of the dark energy parameter,
which in turn is due to improved estimation of the luminosity distance, as shown in Fig.~\ref{Fig:Lumdistz1}. 
Figure~\ref{Fig:Mass_z} depicts how the estimation of $w$
varies with  the total mass of the binary. We show the median errors
from various runs corresponding to different masses and redshifts.
It is interesting to note that for masses between $10^5-10^7M_\odot$,
$\Delta w\leq 3\%$ for redshifts up to 1.5.
The left panel of Fig.~\ref{Fig:Dwerrors} shows that the errors in the measurement of $w$ go 
down by a factor of approximately 10
 when the spins of the binaries are not aligned compared to the aligned, non-precessing case.  
Note that all our histograms, except the cumulative plots in the insets,
are {\it unnormalized}, so that the histograms show the number of
realizations in each bin.
The dependence of the errors in the redshift $z$ is shown in the right panel of Fig.~\ref{Fig:Dwerrors}.  
As is expected the errors get worse with increasing redshift (distance), but the 
degradation of errors is not dramatic.  Even for $z=1.5$, a significant fraction of the $10^4$ 
randomly distributed sources permit measurement of $w$ to better than 5\% accuracy.  
However, for larger redshifts, more galaxies will be found in the LISA error volume 
and it will be more difficult to obtain a redshift. 

Van den Broeck et al.~\cite{VTSS09} have recently revisited the measurement accuracy 
of the dark energy EOS parameter $w$, using a GW signal without precession but including higher harmonics. 
Though higher harmonics and precession are completely different effects, 
it is interesting to note that both improve the estimation of $w$ significantly and to comparable orders. 
This further motivates the attempt to incorporate the effect of higher harmonics in the presence of
spin precession (see \cite{KleinEtAl09} for a recent analysis).  We have
ignored any possible effect of orbital eccentricity in the waveforms 
(see, eg. \cite{YABW09} and references therein).  These are topics for future consideration. 
\acknowledgements
This work was supported in part by the National Science Foundation, 
Grant No.\ PHY 06-52448, the National Aeronautics and Space Administration, Grant No.\
NNG-06GI60G, and the Centre National de la Recherche Scientifique, 
Programme Internationale de la Coop\'eration Scientifique (CNRS-PICS), Grant No. 4396.   

\bibliography{ref-list.bib}

\begin{thebibliography}{32}
\expandafter\ifx\csname natexlab\endcsname\relax\def\natexlab#1{#1}\fi
\expandafter\ifx\csname bibnamefont\endcsname\relax
  \def\bibnamefont#1{#1}\fi
\expandafter\ifx\csname bibfnamefont\endcsname\relax
  \def\bibfnamefont#1{#1}\fi
\expandafter\ifx\csname citenamefont\endcsname\relax
  \def\citenamefont#1{#1}\fi
\expandafter\ifx\csname url\endcsname\relax
  \def\url#1{\texttt{#1}}\fi
\expandafter\ifx\csname urlprefix\endcsname\relax\def\urlprefix{URL }\fi
\providecommand{\bibinfo}[2]{#2}
\providecommand{\eprint}[2][]{\url{#2}}

\bibitem[{\citenamefont{Schutz}(1986)}]{Schutz86}
\bibinfo{author}{\bibfnamefont{B.~F.} \bibnamefont{Schutz}},
  \bibinfo{journal}{Nature (London)} \textbf{\bibinfo{volume}{323}},
  \bibinfo{pages}{310} (\bibinfo{year}{1986}).

\bibitem[{\citenamefont{Markovic}(1993)}]{Markovic93}
\bibinfo{author}{\bibfnamefont{D.}~\bibnamefont{Markovic}},
  \bibinfo{journal}{Phys. Rev. D} \textbf{\bibinfo{volume}{48}},
  \bibinfo{pages}{4738} (\bibinfo{year}{1993}).

\bibitem[{\citenamefont{{Chernoff} and {Finn}}(1993)}]{ChernoffFinn93}
\bibinfo{author}{\bibfnamefont{D.~F.} \bibnamefont{{Chernoff}}}
  \bibnamefont{and} \bibinfo{author}{\bibfnamefont{L.~S.}
  \bibnamefont{{Finn}}}, \bibinfo{journal}{\apj}
  \textbf{\bibinfo{volume}{411}}, \bibinfo{pages}{L5} (\bibinfo{year}{1993}),
  \eprint{arXiv:gr-qc/9304020}.

\bibitem[{\citenamefont{MacLeod and Hogan}(2008)}]{MacLeodHogan07}
\bibinfo{author}{\bibfnamefont{C.~L.} \bibnamefont{MacLeod}} \bibnamefont{and}
  \bibinfo{author}{\bibfnamefont{C.~J.} \bibnamefont{Hogan}},
  \bibinfo{journal}{Phys. Rev. D} \textbf{\bibinfo{volume}{77}},
  \bibinfo{pages}{043512} (\bibinfo{year}{2008}),
  \eprint{arXiv:0712.0618[astro-ph]}.

\bibitem[{\citenamefont{Sathyaprakash et~al.}(2009)\citenamefont{Sathyaprakash,
  Schutz, and Van Den~Broeck}}]{SSV09}
\bibinfo{author}{\bibfnamefont{B.~S.} \bibnamefont{Sathyaprakash}},
  \bibinfo{author}{\bibfnamefont{B.}~\bibnamefont{Schutz}}, \bibnamefont{and}
  \bibinfo{author}{\bibfnamefont{C.}~\bibnamefont{Van Den~Broeck}}
  (\bibinfo{year}{2009}), \eprint{arXiv:0906.4151[astro-ph]}.

\bibitem[{\citenamefont{Cutler and Holz}(2009)}]{CutlerHolz09}
\bibinfo{author}{\bibfnamefont{C.}~\bibnamefont{Cutler}} \bibnamefont{and}
  \bibinfo{author}{\bibfnamefont{D.~E.} \bibnamefont{Holz}}
  (\bibinfo{year}{2009}), \eprint{0906.3752}.

\bibitem[{\citenamefont{Nissanke et~al.}(2009)\citenamefont{Nissanke, Hughes,
  Holz, Dalal, and Sievers}}]{Nissanke09a}
\bibinfo{author}{\bibfnamefont{S.}~\bibnamefont{Nissanke}},
  \bibinfo{author}{\bibfnamefont{S.~A.} \bibnamefont{Hughes}},
  \bibinfo{author}{\bibfnamefont{D.~E.} \bibnamefont{Holz}},
  \bibinfo{author}{\bibfnamefont{N.}~\bibnamefont{Dalal}}, \bibnamefont{and}
  \bibinfo{author}{\bibfnamefont{J.~L.} \bibnamefont{Sievers}}
  (\bibinfo{year}{2009}), \eprint{arXiv:0904.1017[astro-ph.CO]}.

\bibitem[{\citenamefont{Holz and Hughes}(2005)}]{HolzHugh05}
\bibinfo{author}{\bibfnamefont{D.~E.} \bibnamefont{Holz}} \bibnamefont{and}
  \bibinfo{author}{\bibfnamefont{S.~A.} \bibnamefont{Hughes}},
  \bibinfo{journal}{Astrophys.~J} \textbf{\bibinfo{volume}{629}},
  \bibinfo{pages}{15} (\bibinfo{year}{2005}), \eprint{astro-ph/0504616}.

\bibitem[{\citenamefont{Dalal et~al.}(2006)\citenamefont{Dalal, Holz, Hughes,
  and Jain}}]{DaHHJ06}
\bibinfo{author}{\bibfnamefont{N.}~\bibnamefont{Dalal}},
  \bibinfo{author}{\bibfnamefont{D.~E.} \bibnamefont{Holz}},
  \bibinfo{author}{\bibfnamefont{S.~A.} \bibnamefont{Hughes}},
  \bibnamefont{and} \bibinfo{author}{\bibfnamefont{B.}~\bibnamefont{Jain}},
  \bibinfo{journal}{Phys. Rev.} \textbf{\bibinfo{volume}{D74}},
  \bibinfo{pages}{063006} (\bibinfo{year}{2006}), \eprint{astro-ph/0601275}.

\bibitem[{\citenamefont{Apostolatos et~al.}(1994)\citenamefont{Apostolatos,
  Cutler, Sussman, and Thorne}}]{ACST94}
\bibinfo{author}{\bibfnamefont{T.~A.} \bibnamefont{Apostolatos}},
  \bibinfo{author}{\bibfnamefont{C.}~\bibnamefont{Cutler}},
  \bibinfo{author}{\bibfnamefont{G.~J.} \bibnamefont{Sussman}},
  \bibnamefont{and} \bibinfo{author}{\bibfnamefont{K.~S.}
  \bibnamefont{Thorne}}, \bibinfo{journal}{Phys. Rev.~D}
  \textbf{\bibinfo{volume}{49}}, \bibinfo{pages}{6274} (\bibinfo{year}{1994}).

\bibitem[{\citenamefont{Kidder}(1995)}]{K95}
\bibinfo{author}{\bibfnamefont{L.}~\bibnamefont{Kidder}},
  \bibinfo{journal}{Phys. Rev. D} \textbf{\bibinfo{volume}{52}},
  \bibinfo{pages}{821} (\bibinfo{year}{1995}).

\bibitem[{\citenamefont{Vecchio}(2004)}]{Vecchio04}
\bibinfo{author}{\bibfnamefont{A.}~\bibnamefont{Vecchio}},
  \bibinfo{journal}{Phys. Rev. D} \textbf{\bibinfo{volume}{70}},
  \bibinfo{pages}{042001} (\bibinfo{year}{2004}).

\bibitem[{\citenamefont{Lang and Hughes}(2006)}]{LangHughes06}
\bibinfo{author}{\bibfnamefont{R.~N.} \bibnamefont{Lang}} \bibnamefont{and}
  \bibinfo{author}{\bibfnamefont{S.~A.} \bibnamefont{Hughes}},
  \bibinfo{journal}{Phys. Rev.~D} \textbf{\bibinfo{volume}{74}},
  \bibinfo{pages}{122001} (\bibinfo{year}{2006}),
  \bibinfo{note}{erratum-ibid.~{\bf D} 75, 089902 (2007)},
  \eprint{gr-qc/0608062}.

\bibitem[{\citenamefont{Lang and Hughes}(2007)}]{LangHughes07}
\bibinfo{author}{\bibfnamefont{R.~N.} \bibnamefont{Lang}} \bibnamefont{and}
  \bibinfo{author}{\bibfnamefont{S.~A.} \bibnamefont{Hughes}}
  (\bibinfo{year}{2007}), \eprint{arXiv:0710.3795 [astro-ph]}.

\bibitem[{\citenamefont{Stavridis and Will}(2009)}]{SW09}
\bibinfo{author}{\bibfnamefont{A.}~\bibnamefont{Stavridis}} \bibnamefont{and}
  \bibinfo{author}{\bibfnamefont{C.~M.} \bibnamefont{Will}},
  \bibinfo{journal}{Phys. Rev. D} \textbf{\bibinfo{volume}{80}},
  \bibinfo{pages}{040022} (\bibinfo{year}{2009}), \eprint{arXiv:0906.3602}.

\bibitem[{\citenamefont{Yagi and Tanaka}(2009)}]{YagiTanaka09}
\bibinfo{author}{\bibfnamefont{K.}~\bibnamefont{Yagi}} \bibnamefont{and}
  \bibinfo{author}{\bibfnamefont{T.}~\bibnamefont{Tanaka}}
  (\bibinfo{year}{2009}), \eprint{arXiv:0906.4269[gr-qc]}.

\bibitem[{\citenamefont{Lang and Hughes}(2009)}]{LangHughes08}
\bibinfo{author}{\bibfnamefont{R.~N.} \bibnamefont{Lang}} \bibnamefont{and}
  \bibinfo{author}{\bibfnamefont{S.~A.} \bibnamefont{Hughes}},
  \bibinfo{journal}{Class. Quant. Grav.} \textbf{\bibinfo{volume}{26}},
  \bibinfo{pages}{094035} (\bibinfo{year}{2009}),
  \eprint{arXiv:0810.5125[astro-ph]}.

\bibitem[{\citenamefont{Blanchet et~al.}(2006)\citenamefont{Blanchet, Buonanno,
  and Faye}}]{BBuF06}
\bibinfo{author}{\bibfnamefont{L.}~\bibnamefont{Blanchet}},
  \bibinfo{author}{\bibfnamefont{A.}~\bibnamefont{Buonanno}}, \bibnamefont{and}
  \bibinfo{author}{\bibfnamefont{G.}~\bibnamefont{Faye}},
  \bibinfo{journal}{Phys. Rev.~D} \textbf{\bibinfo{volume}{{\bf 74}}},
  \bibinfo{pages}{104034} (\bibinfo{year}{2006}), \bibinfo{note}{erratum-ibid.D
  {\bf 75}, 049903 (E) (2007)}, \eprint{gr-qc/0605140}.

\bibitem[{\citenamefont{Blanchet et~al.}(2004)\citenamefont{Blanchet, Damour,
  Esposito-Far{\`e}se, and Iyer}}]{BDEI04}
\bibinfo{author}{\bibfnamefont{L.}~\bibnamefont{Blanchet}},
  \bibinfo{author}{\bibfnamefont{T.}~\bibnamefont{Damour}},
  \bibinfo{author}{\bibfnamefont{G.}~\bibnamefont{Esposito-Far{\`e}se}},
  \bibnamefont{and} \bibinfo{author}{\bibfnamefont{B.~R.} \bibnamefont{Iyer}},
  \bibinfo{journal}{Phys. Rev. Lett.} \textbf{\bibinfo{volume}{93}},
  \bibinfo{pages}{091101} (\bibinfo{year}{2004}), \eprint{gr-qc/0406012}.

\bibitem[{\citenamefont{Arun et~al.}(2009)\citenamefont{Arun, Buonanno, Faye,
  and Ochsner}}]{ABFO08}
\bibinfo{author}{\bibfnamefont{K.~G.} \bibnamefont{Arun}},
  \bibinfo{author}{\bibfnamefont{A.}~\bibnamefont{Buonanno}},
  \bibinfo{author}{\bibfnamefont{G.}~\bibnamefont{Faye}}, \bibnamefont{and}
  \bibinfo{author}{\bibfnamefont{E.}~\bibnamefont{Ochsner}},
  \bibinfo{journal}{Phys. Rev. D} \textbf{\bibinfo{volume}{79}},
  \bibinfo{pages}{104023} (\bibinfo{year}{2009}),
  \eprint{arXiv:0810.5336[gr-qc]}.

\bibitem[{\citenamefont{Klein et~al.}(2009)\citenamefont{Klein, Jetzer, and
  Sereno}}]{KleinEtAl09}
\bibinfo{author}{\bibfnamefont{A.}~\bibnamefont{Klein}},
  \bibinfo{author}{\bibfnamefont{P.}~\bibnamefont{Jetzer}}, \bibnamefont{and}
  \bibinfo{author}{\bibfnamefont{M.}~\bibnamefont{Sereno}}
  (\bibinfo{year}{2009}), \eprint{arXiv:0907.3318[astro-ph]}.

\bibitem[{\citenamefont{Cutler}(1998)}]{Cutler98}
\bibinfo{author}{\bibfnamefont{C.}~\bibnamefont{Cutler}},
  \bibinfo{journal}{Phys. Rev. D} \textbf{\bibinfo{volume}{57}},
  \bibinfo{pages}{7089} (\bibinfo{year}{1998}).

\bibitem[{\citenamefont{{Berti} et~al.}(2005)\citenamefont{{Berti}, {Buonanno},
  and {Will}}}]{BBW05a}
\bibinfo{author}{\bibfnamefont{E.}~\bibnamefont{{Berti}}},
  \bibinfo{author}{\bibfnamefont{A.}~\bibnamefont{{Buonanno}}},
  \bibnamefont{and} \bibinfo{author}{\bibfnamefont{C.~M.}
  \bibnamefont{{Will}}}, \bibinfo{journal}{Phys.~Rev.~D}
  \textbf{\bibinfo{volume}{71}}, \bibinfo{pages}{084025}
  (\bibinfo{year}{2005}), \eprint{gr-qc/0411129}.

\bibitem[{\citenamefont{Kocsis et~al.}(2006)\citenamefont{Kocsis, Frei, Haiman,
  and Menou}}]{KFHM06}
\bibinfo{author}{\bibfnamefont{B.}~\bibnamefont{Kocsis}},
  \bibinfo{author}{\bibfnamefont{Z.}~\bibnamefont{Frei}},
  \bibinfo{author}{\bibfnamefont{Z.}~\bibnamefont{Haiman}}, \bibnamefont{and}
  \bibinfo{author}{\bibfnamefont{K.}~\bibnamefont{Menou}},
  \bibinfo{journal}{Astrophys. J.} \textbf{\bibinfo{volume}{637}},
  \bibinfo{pages}{27} (\bibinfo{year}{2006}), \eprint{astro-ph/0505394}.

\bibitem[{\citenamefont{{Kocsis} et~al.}(2008)\citenamefont{{Kocsis}, {Haiman},
  and {Menou}}}]{KHMtrig08}
\bibinfo{author}{\bibfnamefont{B.}~\bibnamefont{{Kocsis}}},
  \bibinfo{author}{\bibfnamefont{Z.}~\bibnamefont{{Haiman}}}, \bibnamefont{and}
  \bibinfo{author}{\bibfnamefont{K.}~\bibnamefont{{Menou}}},
  \bibinfo{journal}{\apj} \textbf{\bibinfo{volume}{684}}, \bibinfo{pages}{870}
  (\bibinfo{year}{2008}), \eprint{arXiv:0712.1144}.

\bibitem[{\citenamefont{Arun et~al.}(2007)\citenamefont{Arun, Iyer,
  Sathyaprakash, Sinha, and Van Den~Broeck}}]{AISSV07}
\bibinfo{author}{\bibfnamefont{K.~G.} \bibnamefont{Arun}},
  \bibinfo{author}{\bibfnamefont{B.~R.} \bibnamefont{Iyer}},
  \bibinfo{author}{\bibfnamefont{B.~S.} \bibnamefont{Sathyaprakash}},
  \bibinfo{author}{\bibfnamefont{S.}~\bibnamefont{Sinha}}, \bibnamefont{and}
  \bibinfo{author}{\bibfnamefont{C.}~\bibnamefont{Van Den~Broeck}},
  \bibinfo{journal}{Phys.~Rev.~D} \textbf{\bibinfo{volume}{76}},
  \bibinfo{pages}{104016} (\bibinfo{year}{2007}), \eprint{arXiv:0707.3920
  [astro-ph]}.

\bibitem[{\citenamefont{{Dalal} et~al.}(2003)\citenamefont{{Dalal}, {Holz},
  {Chen}, and {Frieman}}}]{Dalal03WL}
\bibinfo{author}{\bibfnamefont{N.}~\bibnamefont{{Dalal}}},
  \bibinfo{author}{\bibfnamefont{D.~E.} \bibnamefont{{Holz}}},
  \bibinfo{author}{\bibfnamefont{X.}~\bibnamefont{{Chen}}}, \bibnamefont{and}
  \bibinfo{author}{\bibfnamefont{J.~A.} \bibnamefont{{Frieman}}},
  \bibinfo{journal}{\apj} \textbf{\bibinfo{volume}{585}}, \bibinfo{pages}{L11}
  (\bibinfo{year}{2003}), \eprint{arXiv:astro-ph/0206339}.

\bibitem[{\citenamefont{{Pen}}(2004)}]{Pen04WL}
\bibinfo{author}{\bibfnamefont{U.L.} \bibnamefont{{Pen}}},
  \bibinfo{journal}{New Astronomy} \textbf{\bibinfo{volume}{9}},
  \bibinfo{pages}{417} (\bibinfo{year}{2004}), \eprint{arXiv:astro-ph/0305387}.

\bibitem[{\citenamefont{{J{\"o}nsson} et~al.}(2007)\citenamefont{{J{\"o}nsson},
  {Goobar}, and {M{\"o}rtsell}}}]{Goobar06}
\bibinfo{author}{\bibfnamefont{J.}~\bibnamefont{{J{\"o}nsson}}},
  \bibinfo{author}{\bibfnamefont{A.}~\bibnamefont{{Goobar}}}, \bibnamefont{and}
  \bibinfo{author}{\bibfnamefont{E.}~\bibnamefont{{M{\"o}rtsell}}},
  \bibinfo{journal}{\apj} \textbf{\bibinfo{volume}{658}}, \bibinfo{pages}{52}
  (\bibinfo{year}{2007}), \eprint{arXiv:astro-ph/0611334}.

\bibitem[{\citenamefont{{Shapiro} et~al.}(2009)\citenamefont{{Shapiro},
  {Bacon}, {Hendry}, and {Hoyle}}}]{ShapiroEtAl09}
\bibinfo{author}{\bibfnamefont{C.}~\bibnamefont{{Shapiro}}},
  \bibinfo{author}{\bibfnamefont{D.}~\bibnamefont{{Bacon}}},
  \bibinfo{author}{\bibfnamefont{M.}~\bibnamefont{{Hendry}}}, \bibnamefont{and}
  \bibinfo{author}{\bibfnamefont{B.}~\bibnamefont{{Hoyle}}}
  (\bibinfo{year}{2009}), \eprint{{arXiv:0907.3635[astro-ph]}}.

\bibitem[{\citenamefont{Van Den~Broeck et~al.}(2009)\citenamefont{Van
  Den~Broeck, Trias, Sathyaprakash, and Sintes}}]{VTSS09}
\bibinfo{author}{\bibfnamefont{C.}~\bibnamefont{Van Den~Broeck}},
  \bibinfo{author}{\bibfnamefont{M.}~\bibnamefont{Trias}},
  \bibinfo{author}{\bibfnamefont{B.~S.} \bibnamefont{Sathyaprakash}},
  \bibnamefont{and} \bibinfo{author}{\bibfnamefont{A.}~\bibnamefont{Sintes}},
  \emph{\bibinfo{title}{Measuring the dark energy equation of state with \emph{LISA}}}
  (\bibinfo{year}{2009}), \bibinfo{note}{poster presented at the 8th Edoardo
  Amaldi Conference on Gravitational Waves, New York}.

\bibitem[{\citenamefont{Yunes et~al.}(2009)\citenamefont{Yunes, Arun, Berti,
  and Will}}]{YABW09}
\bibinfo{author}{\bibfnamefont{N.}~\bibnamefont{Yunes}},
  \bibinfo{author}{\bibfnamefont{K.~G.} \bibnamefont{Arun}},
  \bibinfo{author}{\bibfnamefont{E.}~\bibnamefont{Berti}}, \bibnamefont{and}
  \bibinfo{author}{\bibfnamefont{C.~M.} \bibnamefont{Will}}
  (\bibinfo{year}{2009}), \eprint{arXiv:0906.0313.}

\end{thebibliography}
\end{document}